\magnification=1200 \vsize=25truecm \hsize=16truecm \baselineskip=0.6truecm
\parindent=1truecm \nopagenumbers \font\scap=cmcsc10 \hfuzz=0.8truecm
\font\tenmsb=msbm10
\font\sevenmsb=msbm7
\font\fivemsb=msbm5
\newfam\msbfam
\textfont\msbfam=\tenmsb
\scriptfont\msbfam=\sevenmsb
\scriptscriptfont\msbfam=\fivemsb
\def\Bbb#1{{\fam\msbfam\relax#1}}

\null \bigskip  \centerline{\bf CONSTRUCTING INTEGRABLE THIRD ORDER SYSTEMS:}
\centerline{\bf THE GAMBIER APPROACH}

\vskip 2truecm
\bigskip
\centerline{\scap S. Lafortune$^{\dag}$}
\centerline{\sl LPTM et GMPIB,  Universit\'e Paris VII}
\centerline{\sl Tour 24-14, 5$^e$\'etage}
\centerline{\sl 75251 Paris, France}
\footline{\sl $^{\dag}$ Permanent address: CRM, Universit\'e de
Montr\'eal, Montr\'eal, H3C 3J7 Canada}

\bigskip
\centerline{\scap B. Grammaticos}
\centerline{\sl GMPIB (ex LPN), Universit\'e Paris VII}
\centerline{\sl Tour 24-14, 5$^e$\'etage}
\centerline{\sl 75251 Paris, France}
\bigskip
\centerline{\scap A. Ramani}
\centerline{\sl CPT, Ecole Polytechnique}
\centerline{\sl CNRS, UPR 14}
\centerline{\sl 91128 Palaiseau, France}

\vskip 2truecm \noindent Abstract \medskip
\noindent  We present a systematic construction of integrable third order
systems based on
the coupling of an integrable second order equation and a Riccati equation.
This approach is
the extension of the Gambier method that led to the equation that bears his
name. Our study is
carried through for both continuous and discrete systems. In both cases the
investigation is
based on the study of the singularities of the system (the Painlev\'e
method for ODE's and
the singularity confinement method for mappings).

\vfill\eject

\footline={\hfill\folio} \pageno=2

\bigskip
\noindent {\scap 1. Introduction}
\medskip
\noindent The investigation of the integrability of second order
differential equations has
been one of the most important enterprises in the history of integrable
systems. Initiated
by Painlev\'e [1] and completed by Gambier [2], it established the
importance of singularity
analysis as an integrability criterion. The results of the  Painlev\'e-Gambier
investigations are of capital importance since they showed the existence of new
transcendents,  known since then under  the name of  Painlev\'e.
Overshadowed by
this momentous discovery, the work of Gambier on linearizable systems did
not receive the
attention it deserved. The recent discovery of integrable {\sl discrete}
systems has
led naturally to a critical examination of the work of the $19^{\rm th}$
century masters. In
particular, we have shown that it is possible to find discrete forms not
only for the
Painlev\'e equations, but, in fact, for every single equation in the
Painlev\'e-Gambier
list. The equation \#XXVII of the list of 50 canonical equations [3], which
we decided to
call the Gambier equation, was of course among them. Its discretization
necessitated a
thorough understanding of the Gambier approach.

The key idea of Gambier (but we are conscious that the historical truth may be
different) was to construct an integrable
second order equation by suitably coupling two integrable first order ones. The
latter were well-known: at first order the only integrable ordinary
differential equations
are either linear or of Riccati type. The Gambier equation is precisely the
coupling of two Riccati in cascade (and it contains as a subcase the
coupling involving one
or even two linear equations). From the point of view of singularity
analysis this coupling
of two integrable equations is not harmless. Each of the equations has
singlevalued movable
singularities. However, the singularities induced on the second equation by the
singularities of the solution of the first one (and which would thus look
superficially as
fixed) may lead to multivaluedness. This feature makes the
application of singularity analysis mandatory. Its implementation leads to the
(algebraically) integrable forms of the Gambier equation.

In perfect analogy to the continuous case, we have introduced in [4] the
Gambier
mapping. The latter is a system of two coupled homographic mappings (which
play the role of
the discrete Riccati) in cascade. The integrable forms were obtained
through the application
of the discrete integrability criterion that we have proposed under the
name of singularity
confinement.

In the present work we shall address the question of the construction of
integrable
third order systems in the spirit of Gambier. Namely we shall start with a
second order integrable equation and couple it with a Riccati (or a linear)
first order
(also integrable) equation. This enterprise may easily assume staggering
proportions. While
at second order one had only two first order building blocks at one's
disposal, at third
order there are minimally 24 equations (the Gambier list) to be coupled to
the two first order integrable ones. The situation is even more
overwhelming in the discrete case
since it is well-known that each continuous equation of the Gambier list
may possess several
discrete avatars. In order to limit the scope of our investigation we shall
consider coupled systems where the dependent variable enters only in a
polynomial way. This
leads naturally to the coupling of a Painlev\'e ($\Bbb P$) I or II to a
Riccati.

Historically the coupling of a $\Bbb P$ equation with a Riccati was first
considered by
Chazy [5].  He examined an additive coupling of P$_{\rm I}$ with a Riccati.
Starting with
P$_{\rm I}$ in the form:
$$w''=6w^2+z,$$
he introduced a Riccati:
$$y'=\alpha y^2+\beta y +\lambda w+\gamma. \eqno(1.1)$$
This coupling is additive as opposed to the one introduced by Gambier which
is multiplicative
and assumes the form:
$$y'=\alpha y^2+(\beta + \lambda w)y + \gamma.$$
(In the case of the Gambier coupling $w$ is the solution of a Riccati
equation).
Since the singularities of  P$_{\rm I}$ are double poles
\big(${6/(z-z_0)^2}$\big), the
only coupling that is compatible with integrability is the additive one.
Assuming that
$\alpha \ne 0$, we can put $\beta =0$ by a simple translation of $y$ and
Chazy found that the
only cases where the leading singularity does not induce multivaluedness
were when equation
(1.1) assumed the form:
$$y'={1-k^2 \over 4}y^2 + w + \gamma, \eqno(1.2)$$
where $k$ is an integer not multiple of 6. Thus 5 cases had to be examined
as $k=6m+n$ with
$n=1,\ldots,5$. Chazy found the following necessary integrability constraints:
$$
\matrix{
n=2 &\gamma    =0  \hfill\cr
n=3 &\gamma'  =0   \hfill\cr
n=4 &\gamma''  =\mu\gamma^2+\nu z  \hfill \cr
n=5 &\gamma''' =\mu\gamma\gamma' + \nu,\hfill \cr
}$$
where $\mu$ and $\nu$ are specific numerical constants. It turns out that
for $k=n$ they
are also sufficient. For
$k=6m+1$ the first condition appears at $k=7$. In this case the constraint
reads:
$$
\gamma^{(5)}=48 \gamma \gamma ''' +120 \gamma'\gamma''-{2304
\over 5}\gamma' \gamma^2 - 24z\gamma'-48\gamma.
$$
This equation has the $\Bbb P$ property and is thus expected to be
integrable. Still it is
interesting to point out that this equation is more difficult to solve than
the one we
started with which is of third order.

Chazy offers only a rapid comment concerning the case $k \geq 8$. In fact,
the constraints
obtained are necessary, but not sufficient for higher $k$'s. We have
examined the
first few cases beyond $k=7$ using the same method as Chazy, namely
singularity analysis
(but unlikely Chazy our approach has profited from the existence of
computer algebra tools).
It turned out that none of the cases we examined satisfied the Painlev\'e
criterion. So,
although this is not a proof in a strict sense, we can suppose that no
integrable cases exist
beyond the 5 identified by Chazy.

\bigskip
\noindent {\scap 2. Coupling of integrable second order ODE's with a Riccati}
\medskip
\noindent

As we have explained in the introduction we shall not attempt an exhaustive
treatment of all
24 [2] (or 50 [3], or more [6]) second order equations of the
Painlev\'e/Gambier list with a
Riccati. Instead we shall limit ourselves to the simplest case, namely
equations where the
dependent variable enters in a polynomial way (instead of rational). This
limits the
research to just three generic equations: P$_{\rm I}$ (already examined by
Chazy), P$_{\rm
II}$ and the linearisable, G5 (number 5 of the Gambier list), equation.
Both P$_{\rm I}$
and G5 have dominant singularities that are single poles i.e. $w \sim
1/(z-z_0)$. Thus the
adequate coupling is through a multiplicative Riccati. An additive coupling
would lead to
logarithmic singularities in the Riccati and thus to multivaluedness
incompatible with
integrability.

\bigskip
\noindent {\scap {\sl 2a  Coupling P$_{\rm II}$ with a Riccati}}
\medskip
\noindent

We start with the canonical form of P$_{\rm II}$, namely:
$$
w''=2w^3+zw+\mu \eqno(2.1)
$$
and consider the following multiplicative coupling:
$$
y'=\alpha y^2+(nw+\beta)y + \gamma. \eqno(2.2)
$$
 A gauge transformation on $y$ can be used in order to put $\beta$
to zero. Next, we proceed to determine $\alpha, \gamma$ through the
application of
singularity analysis so as to ensure the Painlev\'e property for the
system. Equation (2.1)
has of course the $\Bbb P$ property, and the expansion of its solution
around a singularity
is:
$$
w={\sigma \over z-z_0} + \dots + a_4(z-z_0)^4 + \dots \eqno(2.3)
$$
where $\sigma ^2=1$ and $a_4$ is a free parameter (the second one beside
$z_0$). The
coupling of $w$ with $y$ must not lead to multivaluedness. Thus the
coefficient $n$ of the
coupling must be an integer. This is only the first condition and, by far, not
sufficient. In order to proceed further we expand $y$ around the
singularity $z_0$ and
assume that $y$ either has a pole at the same location or it is regular.
Substituting the
expansion form (2.3) we can compute the terms of the series of $y$ and
obtain the
compatibility conditions for the absence of logarithmic terms in the
expansion of $y$. We
find thus the following condition for $n=1$:
$$
\gamma=\alpha    =0.
\eqno(2.4)$$
Let us point out that $\alpha=\gamma=0$ works for every value of $n$: $w$
is simply related to
the logarithmic derivative of $y$. We have furthermore, for $n=2$:
$$
\gamma'=\alpha'    =0. \eqno(2.5)
$$
For $n=3$ we obtain
$$
{\gamma \over \alpha}=-\left({2\alpha''+\alpha z \over \alpha^3}\right)
{\rm and}\;\;
{\alpha \over \gamma}=-\left({2\gamma''+\gamma z \over \gamma^3}\right).
\eqno(2.6)
$$
Isolating $\gamma$ and integrating once, we find:
$$
\alpha\alpha'''-3\alpha'\alpha''-\alpha'\alpha z-k\alpha^2=0 \eqno(2.7)
$$
and putting $\phi={\alpha' \over \alpha}$ we find
$$
\phi''=2\phi ^3+z\phi +k. \eqno(2.8)
$$
Thus the logarithmic derivative of $\alpha$ satisfies precisely P$_{\rm
II}$ (with a free
constant $k$).
For $n=4$ we find a more complicated condition:
$$
3\alpha \gamma \gamma' +
\alpha'\gamma^2+3\alpha(1+\mu)+3\gamma'z+9\gamma'''=0, \eqno(2.9a)
$$
$$
3\gamma \alpha \alpha' + \gamma'\alpha^2+3\gamma(1-\mu)+3\alpha'z+9\alpha'''=0.
\eqno(2.9b)$$
Putting $\alpha^2=\phi'$ we can integrate (2.9b) (multiplied by $\alpha$)
for the quantity
$\alpha^3\gamma$ and thus obtain $\gamma$. Then (2.9a) gives a 6$^{\rm th}$
order homogeneous
equation for $\phi$ and putting $u=\phi'/\phi$ leads to a 5$^{\rm th}$
order equation. This
equation passes the $\Bbb P$ test and is thus presumably integrable but its
integration
is a more complicated task than the equation we started with, which is only
of third order.

For $n=5$ we obtain again as a first condition $\alpha= \gamma=0$ which as
we explained is
sufficient. For $n=6$ a first condition (as in the $n=2$ case) is
$\alpha'=\gamma'=0$. However
a second condition appears. In fact, for $n>4$ the free parameter of the
expression (2.3)
$a_4$ starts appearing in the compatibility condition which must be
identically satisfied.
Thus for
$n=6$ we find the second the second condition either $\alpha=0$ and
$\mu=-7/6$ or $\gamma=0$
and $\mu=7/6$. Thus this coupling works only for some particular case of
P$_{\rm II}$ with a
specific $\mu$. For $n \geq 7$ we have not been able to find any integrable
case, besides
the trivial $\alpha=\gamma=0$ one. In some cases it is even possible to
prove the
incompatibility of the constraints. We surmise that the multiplicative
coupling of P$_{\rm
II}$ with a Riccati does not possess any integrable case besides the ones
listed above.

\bigskip
\noindent {\scap {\sl 2b Coupling the linearisable G5 with a Riccati}}
\medskip
\noindent
The canonical form of the linearisable equation, G5 in the Gambier list is:
$$
w''=-3w'w-w^3+q(z)(w'+w^2). \eqno(2.10)
$$
The Cole-Hopf transformation $w=-u'/u$ reduces (2.10) to a linear equation
$u'''=q(z)u''$.
The function $q(z)$ is completely free. Given this fact one can make two
different couplings.
The first is the `standard' one where the solution of (2.10) for given
$q(z)$ is injected
into a Riccati:
$$
y'=\alpha y^2+nwy+\gamma. \eqno(2.11)
$$
The condition for Painlev\'e property for $n<0$ turns out to be $\alpha=0$,
while for $n>0$
it is $\gamma=0$. In both cases $(2.10)$ becomes a linear equation (either
for $y$ or for
$1/y$)  and the remaining free function ($\gamma$ or $\alpha$) does not produce
multivaluedness.

The second case of coupling is when $q(z)$  is itself proportional to the
solution of a
Riccati. Thus the coupled system now becomes
$$
\eqalignno{
&w'=\alpha w^2+\beta w+\gamma, & (2.12a) \cr
&y''=-3yy'-y^3+nw(y'+y^2). & (2.12b)
}$$
Only the case $\alpha\neq 0$ needs to be considered since when $\alpha=0$
equation (2.12a) is
linear and thus does not have any singularities. Since $\alpha\neq 0$ we
can take $\alpha=1$.
As previously the coupling enters through $nw$ with integer $n$ since the
singularity of
(2.12a) is a simple pole. For $n<0$ the system has always the Painlev\'e
property and thus
$\beta$ and $\gamma$ are free. On the contrary, for $n>0$ we have stringent
integrability
conditions. For
$n=1,2$ there is no solution for $\beta$, $\gamma$ leading to the
Painlev\'e property for the
system. For $n=3$ we find as only solution $\beta = \gamma =0$. For $n=4$
we obtain the
condition:
$$
\gamma=-{11 \over 4} \beta^2 + \beta' \eqno(2.13a)
$$
and
$$
\beta''=12\beta\beta' -16 \beta^3 \eqno(2.13b).
$$
Putting $\beta =-{\phi ' \over 4\phi}$, (2.13b) reduces to $\phi'''=0$ and
we thus have
elementary expressions for $\beta$ and $\gamma$.

For $n\geq 5$ we can obtain the two compatibility conditions in the form of
a higher order
nonlinear system for $\beta, \gamma$. It turns out that for the first few
cases studied this
system has the weak ${\Bbb P}$ property [7]. We have not tried to integrate
these systems since
their integration is more difficult than the problem we started with.

\bigskip
\noindent {\scap 3. Coupling of a second order mapping with a discrete
Riccati equation}
\medskip
\noindent

Constructing integrable discrete systems in the same spirit as Gambier is quite
straightforward once the basic ingredients are available. What is needed is
a detailed
knowledge of the forms of the equations to be coupled and a reliable
integrability
detector. The second order mappings which play the role of the ${\Bbb P}$
equations in the
discrete domain have been the object of numerous detailed studies and we
are now in possesion
of discrete forms of all the equations of the Painlev\'e/Gambier
classification. The discrete
integrability detector is based on the singularity confinement that we
discovered in [8] and
which has turned out to be of uptmost reliability.

The coupling we are going to consider is a homographic mapping (discrete
Riccati) for the
variable $y$:
$$
\overline y = {(\alpha x + \beta)y +(\eta x + \theta) \over (\epsilon x +
\zeta)y +(\gamma x +
\delta)}, \eqno(3.1)
$$
(where $\overline y$ stands for $y_{n+1}$, $y$ for $y_n$ and $\alpha$, $\beta$,
$\dots$,$\theta$ depend in general on $n$) the coefficients of which depend
linearly on $x$,
the solution of the discrete ${\Bbb P}$ I or II. (We shall not present here
the coupling of the
discrete analog of the linearisable equation to a Riccati. As a matter of
fact this equation
is the simplest non-trivial member of the hierarchy of projective Riccati
systems, the
discretisation of which was presented in full generality in [9]). The
mapping (3.1) can be
simplified and brought under canonical form through the application of
homographic
transformations on $y$. The generic form of the result is:
$$
\overline y = {(\alpha x + \beta)y +1 \over y +(\gamma x +
\delta)}. \eqno(3.2)
$$
Non-generic cases do exist as well, and foremost among those is the linear
relation:
$$
\overline y (\gamma x +\delta)-y(\alpha x + \beta)-1=0. \eqno(3.3)
$$
In what follows we shall examine in detail the coupling of (3.2) and (3.3)
with either
d-P$_{\rm I}$ or d-P$_{\rm II}$ (under various forms).

How does one apply the singularity confinement criterion to a mapping such
as (3.2) when $x$
is given by some discrete equation like d-P$_{\rm I}$ or d-P$_{\rm II}$?
The first step
consists in determining the singularities of (3.2). As we have explained in
[10] the
singularity manifests itself by the fact that $\overline y$ is independent
of $y$. (We say in this
case that $y$ ``forgets the initial condition'' or ``loses one degree of
freedom''). The
condition for $\overline y$ to be independent of $y$ is just
$$
(\gamma x + \delta)(\alpha x + \beta)=1. \eqno(3.4)
$$
This quadratic equation has two roots which we will denote by $X_1$, $X_2$:
they can be
easily related to $\alpha$, $\beta$, $\gamma$, $\delta$. The confinement
condition is for $y$
to recover the lost degree of freedom. This can be done if $y$ assumes an
indeterminate form
$0/0$. This means that $x$ at this stage must again satisfy (3.4) and
moreover be
such that the denominator (or, equivalently, the numerator) vanishes.

Let us assume now that for some $n$ we have $x_n=X_1$. The confinement
requirement is that $k$
steps later $x_{n+k}=X_2$. (We must point out here that we require that
$x_{n+k}$ be equal
to $X_2$ and not to $X_1$ again. If the latter were done this would mean
that $X_1$ is a
singularity occuring periodically. Such singularities are not really
movable, i.e. their
position cannot be freely adjusted by choosing the appropriated initial
conditions. Our
conjecture is that they do not play any role in integrability, just like
the fixed
singularities in the continuous case). Starting from $x_n=X_1$ and some
initial datum
$x_{n-1}$, we can iterate the mapping for $x$ and obtain $x_{n+k}$ as a
complicated function
of $x_{n-1}$ and $X_1$. Since
$x_{n+k}$ depends on the free parameter $x_{n-1}$ there is no hope for
$x_{n+k}$ to be
equal to $X_2$ if $X_1$ is a generic point for the mapping of $x$.
The only possibility is
that both
$X_1$ and
$X_2$ be special values. What are the special values of this equation
depends on its details, but clearly in the case of the discrete ${\Bbb
P}$'s we shall examine
here, these values can only be the ones  related to the singularities. To
be more
specific, let us examine d-P$_{\rm II}$:
$$
\overline x + \underline x = {zx+a \over 1-x^2}. \eqno(3.5)
$$
The only special values of $x$ are the ones related to the singularity
$x_n=\pm 1$,
$x_{n+1}=\infty$, $x_{n+2}=\mp 1$ while $\dots$, $x_{n-2}$, $x_{n-1}$ and
$x_{n+3}$, $x_{n+4}$,
$\dots$ are finite. This means that the two roots of (3.4) must be two of
$\{+1, \infty,
-1\}$  and moreover that confinement must occur in two steps. The precise
implementation of singularity confinement requires that the denominator of
(3.2) at $n+2$
vanishes (and because of (3.4) this ensures that the numerator vanishes as
well). Moreover,
we must make sure that the lost degree of freedom (i.e. the dependence on
$y$) is indeed
recovered through the indeterminate form.

\bigskip
\noindent {\scap {\sl 3a Coupling various d-P$_{\rm I}$'s with a discrete
Riccati}}
\medskip
\noindent

In this subsection we are going to analyse the coupling of four different
forms of d-P$_{\rm
I}$ to the homographic mapping (3.2) and to a linear equation (3.3). The
d-P$_{\rm I}$'s we
are going to consider are the following (presented below together with
their singularity
pattern):
$$
\eqalignno{
&\overline x + \underline x = {z_n \over x}+{1 \over x^2}\hskip
1.39cm\{0,\infty,0\},& (3.6)
\cr
&\overline x + x + \underline x = {z_n \over x}+1\hskip
1.03cm\{0,\infty,\infty,0\},& (3.7)
\cr
&\overline x + \underline x = {z_n \over x}+1 \hskip
1.66cm\{0,\infty,1,\infty,0\},& (3.8)
\cr
&\overline x  \underline x = -{q_n \over x^2}+{1\over x}\hskip
1.75cm\{q,0,\infty,0,q\},&
(3.9) }$$
with $z_n=an+b$ and $q_n=q_0\lambda^n$ ($a$, $b$, $q_0$, $\lambda$ are
constants). (More forms
of the discrete P$_{\rm I}$ [11] are known but we shall restrict our
analysis of the possible
couplings to just these simplest forms).

All the singularity patterns above have as common characteristic that one
can enter the
singularity through $0$ and exit it again through $0$. This means that the
condition (3.4)
can have $0$ as a double pole. This results to the following conditions:
$$
\matrix{
\displaystyle{\beta \delta =1}, \hfill \cr
\displaystyle{\alpha \delta + \beta \gamma=0},
} \eqno(3.10)
$$
and since neither $\alpha$ nor $\gamma$ can vanish (lest the $x^2$ term
disappear) we have
$\delta=1/\beta$ and $\gamma=-\alpha/\beta^2$. One can, of course, consider
the case where one
(or two) of the roots of (3.4) are equal to $\infty$: after all $\infty$ is
part of the
special values of the singularity pattern. It has turned out that except
for the case (3.8)
the consideration of these cases does not lead to any interesting result.
(Let us point out
that the value $1$ appearing in the singularity pattern of (3.8) should not
be considered as a
special value: it may well occur outside any singularity pattern). Thus the
first discrete
Riccati we are going to consider is of the form:
$$
\overline y = {y(\alpha x + \beta)+1 \over y-(\alpha x -\beta)/ \beta^2}.
$$
In all the cases considered, the first confinement condition, namely that
$y$ (at a suitable
$n$) assumes the form $0/0$ does not suffice in order to reintroduce the
dependence in the
initial conditions.  It is thus necessary to proceed to the next order and
introduce one
further constraint (which turns out to be sufficient). Let us work out in
detail the
case of the d-P$_{\rm I}$ (3.6). Starting with $\underline x=0$ we obtain
$y=\underline \beta$
i.e. independent of the value of $\underline y$. For $x=\infty$ we obtain
$\overline y =
-\underline \beta \beta^2$ and finally at the next step, $\overline x =0$,
we ask that the
numerator and denominator of $\overline{\overline y}$ vanish. This leads to
the first condition
$$
\underline \beta \beta^2 \overline \beta =1. \eqno(3.12)
$$
Implementing this constraint leads to a second confinement condition that
reads: $\underline
\alpha / \underline \beta=\overline \alpha / \overline \beta$. This means
that $\alpha = c\beta$
where $c$ is a constant with an even-odd dependence. The solution of the
constraint (3.12) is
straightforward. Taking the logarithm of both members and calling
$b=\log{\beta}$ we find the
linear equation
$$
\underline b + 2b + \overline b =0 \eqno(3.13)
$$
with solution $b=(p+qn)(-1)^n$. Simple solutions to (3.12) can be obtained
from this last
solution. On the other hand just by inspection we can obtain solutions to
(3.12) where
$\beta$ is constant: $\beta = \pm 1, \pm i$.

The case of the ``standard'' d-P$_{\rm I}$ (3.7) can be treated along
similar lines. The first
confinement condition reads:
$$
\underline \beta \beta^2 {\overline \beta}^2
\overline{\overline{\beta}}=-1
\eqno(3.14)
$$
while the second becomes too complicated to be exactly solved. We prefer to
proceed using one
particular solution of (3.14) corresponding to constant $\beta$'s, for
example $\beta = i$.
This leads to a second confinement condition $\underline \alpha / \underline
z=\overline{\overline \alpha}/\overline{\overline z}$. Thus $\alpha = cz$
where $c$ is a
constant with ternary freedom ($\overline{\overline c} = \underline c$). If
we implement
$\beta =e^{\pm i\pi /6}$ and define $\chi=-\alpha z$, we get, as a second
confinement
condition, the equation:
$$
\overline \chi + \chi + \underline \chi = 3 \beta^2{z \over \chi} + c,
\eqno(3.15)
$$
where $c$ is a constant of integration. Thus, after considering the
coupling with a d-P$_{\rm
I}$ we get a d-P$_{\rm I}$ of the same type as one of the confinement
conditions. This is
in perfect parallel to the continuous case of Chazy (coupling (1.2) with
$n=4$) where we get
another P$_{\rm I}$ as the integrability condition for a coupling between a
Riccati and a P$_{\rm
I}$.

 The case (3.8) leads to
still more complicated equations. One way to simplify them is to choose
$\beta$ satisfying:
$$
{\underline  \beta}\, \beta^2 {\overline \beta}
=1  \eqno(3.16)
$$
which is sufficient (but not necessary) to satisfy the first confinement
condition.
We can then implement the solutions $\beta=i$ and $\beta=1$. If $\beta=i$,
the second
confinement condition is $\alpha=c/z$ (where c is a constant with
quaternary freedom
$\underline{\underline c}=\overline{\overline c}$). If
$\beta=1$, we define $\chi=-\alpha z$ and we get, as second confinement
condition, the following
equation:
$$
\overline \chi + \underline \chi = -{4z \over \chi} + c, \eqno(3.17)
$$
where $c$ is a constant with binary freedom. So, again here, we get a
d-P$_{\rm I}$ of the
same type as the one we started with as a confinement condition.

For the case (3.8), it is also possible to consider a coupling where the
condition (3.4) has
$0$ and $\infty$ as roots. This means that $\alpha=0$ (we could also choose
$\gamma=0$ but
these two cases are equivalent under the homographic transformation
$w\rightarrow 1/w$) and
$\delta=1/\beta$. The first confinement condition then is $\gamma=-\delta$.
We define
$\chi=\overline \beta \beta$ and the second integrability condition reads:
$$
\overline \chi + \underline \chi =-{\underline z + c \over \chi} + 1,
\eqno(3.18)
$$
where $c$ is a constant of integration. Thus again we get a d-P$_{\rm I}$
of the
same type as the one we started with. Finally we can also consider the case
where the
condition (3.4) has $\infty$ as a double root. We then must have
$\alpha=\beta=0$. The first
confinement condition is $\delta=-\gamma$ and  we
get the following relation for $\gamma$:
$$
\overline \gamma \gamma={1 \over -\underline z + k}, \eqno(3.19)
$$
(where $k$ is a constant of integration) which can be solved in an
elementary way for $\gamma$.

In the case of $q$-P$_{\rm I}$ (3.9) the full singularity pattern is one
where we enter the
singularity at $q$ and exit it at $q$ after four steps. However the
complete study of this
singularity pattern turns out to be untractable. Thus we shall limit
ourselves here to the case
where we enter the singularity through $0$ and exit it through $0$ after
two steps. In this case
the first confinement condition is just (3.12). Once this is implemented
the second condition
reads
$\overline \alpha \underline \beta = \lambda \underline \alpha \overline
\beta$ which means
$\alpha = c\beta\mu^n$ where $c$ is a constant with binary freedom and
$\mu^2=\lambda$.

Let us now turn to the case of the coupling of d-P$_{\rm I}$ with a linear
equation (3.3).
For the special values of d-P$_{\rm I}$ $0$ and $\infty$, only three
couplings have to
be considered:
$$
\eqalignno{
&\overline y = \alpha y + {1 \over \gamma x}, & (3.20a) \cr
&\overline y = \alpha xy + {1 \over \delta}, & (3.20b) \cr
&\overline y = {\beta y + 1 \over \gamma x}. & (3.20c)
}
$$
It turns out that in every case examined the second (3.20b) and third
(3.20c) are always
incompatible with confinement. The only remaining candidate is thus the
coupling of the form
(3.20a). By the appropriate gauge of $y$ we can bring it to the form:
$$
\overline y - y = {1 \over \gamma x}. \eqno(3.21)
$$
Let us work out in detail the case of (3.6). A detailed analysis of the
singularity pattern
shows that if $\underline x$ vanishes like $\epsilon$ then $x$ diverges
like $1/\epsilon^2$
ans $\overline x$ vanishes like $-\epsilon$. We compute the corresponding
$y$'s and find, at
leading order, $y\sim 1/\underline \gamma \epsilon$, $\overline y\sim
1/\underline \gamma
\epsilon$ and the condition for $\overline{\overline y}$ to be finite is
$1/\underline \gamma
-1/\overline \gamma = 0$ i.e. $\gamma$ must be a constant with binary
freedom (i.e.
even-odd dependence). The analysis of the remaining cases proceeds along
similar lines. For
(3.7) we have the pattern $\{\epsilon, \underline z/\epsilon,-\underline
z/\epsilon, -\epsilon
\overline{\overline z}/\underline z \}$ and the condition for
$\overline{\overline {\overline
y}}$ to be finite is $\overline{\overline \gamma}\, \overline{\overline z}
= \underline \gamma
\underline z$ i.e. $\gamma=k/z$ where $k$ is a constant with ternary
freedom.  The case (3.8) is
related to the pattern $\{\epsilon,
\underline{\underline z}/\epsilon,1,-\underline{\underline
z}/\epsilon,\overline{\overline
z} \epsilon/\underline{\underline z} \}$ leading to the confinement condition
$\underline{\underline \gamma}\,\underline{\underline z}=
\overline{\overline \gamma}\,
\overline{\overline z}\;$ i.e. $\gamma=k/z$ where $k$ is a constant with
quaternary freedom.
Finally the case (3.9) is related to the pattern $\{\underline{\underline
q}+\epsilon,
a\epsilon,-\lambda/(a^2\epsilon^2), -\epsilon a/\lambda,
\overline{\overline q}\}$ (where $a$ is
a free constant). Again we concentrate on the singularity induced by $x=0$
and which confines when
$x=0$ again. This results to the condition
$\overline
\gamma =
\lambda
\underline \gamma$ which means $\gamma=k\mu^n$ where $k$ is a constant with
binary freedom and
$\mu^2=\lambda$.

\bigskip
\noindent {\scap {\sl 3b Coupling discrete P$_{\rm II}$'s with a discrete
Riccati}}
\medskip
\noindent

In this section we shall examine the coupling of two different discrete
forms of P$_{\rm II}$
with a Riccati: a difference one (which is the ``standard'' d-P$_{\rm II}$)
$$
\overline x + \underline x = {zx+\mu \over 1-x^2}, \eqno(3.22)
$$
where $z$ is linear in the discrete variable $n$ and $\mu$ is a constant,
and one of $q$-type:
$$
\overline x \underline x = {p(x-q) \over x(x-1)}, \eqno(3.23)
$$
where $q=q_0\lambda^n$ and $p=p_0\lambda^{n}$.

Let us start with d-P$_{\rm II}$ (3.22). The singularity pattern of this
equation is $\{\pm
1,\infty,\mp 1\}$. This means that the singularity condition (3.4) must
have $\pm 1$ as roots
(the case when one root is $\infty$ does not lead to anything interesting).
As a result we
have that $\delta$ and $\gamma$ are given by
$\delta=-\beta/(\alpha^2-\beta^2)$,
$\gamma=\alpha/(\alpha^2-\beta^2)$. The pattern $\{1,\infty,-1\}$ leads to
a confinement
condition: $\gamma=(\underline \alpha + \underline \beta)\alpha(\overline
\alpha - \overline
\beta )$ while the second pattern $\{-1,\infty,1\}$ leads to
$\gamma=(\underline \alpha -
 \underline \beta)\alpha(\overline \alpha + \overline \beta )$. Equating
the two expressions
for $\gamma$ we find $\underline \alpha \overline \beta = \overline \alpha
\underline \beta$ i.e. $\beta=k\alpha$ where $k$ is a constant with binary
freedom which we
will ignore from now on. Expressing $\gamma$ in two possible ways we get
finally for $\alpha$
the equation:
$$
\underline \alpha \alpha^2 \overline \alpha = {1 \over (1-k^2)^2}. \eqno(3.24)
$$
This equation can be solved by linearisation just by taking the logarithm
of both sides.

The $q$-P$_{\rm II}$ has also two singularity patterns $\{q,0,\infty,1\}$ and
$\{1,\infty,0,q\}$. Requiring $0$ and $1$
to be roots of (3.4) gives the following expressions for
$\gamma$, $\delta$: $\delta=1/\beta$ and $\gamma = -{\alpha \over
\beta(\alpha + \beta)}$.
Next we obtain the confinement conditions for the two patterns of
singularities:
$$
\eqalignno{
&\underline \alpha \alpha \beta \overline \beta +\alpha \underline \beta
\beta	\overline
\beta + \underline \alpha \beta^2\overline \beta + \underline \beta \beta^2
\overline \beta
-1=0, & (3.25a) \cr
& \alpha \overline \alpha \underline \beta \beta +\overline \alpha
\underline \beta \beta^2
 + \alpha \underline \beta \beta \overline \beta + \underline \beta \beta^2
\overline \beta
-1=0. & (3.25b)
}
$$
Subtracting these two equations we obtain $\overline \alpha \underline
\beta = \underline
\alpha \overline \beta$ i.e. $\beta=k\alpha$ where $k$ is a constant with
binary freedom
which we again ignore. Substituting back to (3.25) we obtain the final
condition:
$$
\underline \alpha \alpha^2 \overline \alpha = {1 \over k^2(k+1)^2} \eqno(3.26)
$$
which can be integrated through linearisation as explained above. The case
where (3.3)
has $\infty$ and
$q$ as roots is equivalent to the one treated above by a homographic
transformation.

We now consider the case where (3.4) has $0$ and $q$ as roots which impose
the relations
$\delta=1/\beta$ and $\gamma={-\alpha \over \beta ( q\alpha+\beta)}$. As
first condition
we then get that $\beta$ is a constant with binary freedom. We ignore this
freedom and
consider $\beta$ as constant and we get the following relation for $\alpha$:
$\alpha=-{(\beta^2+1) \over \beta q}$. Finally, the case where (3.4) has
$q$ and
$1$ as roots has been studied but the resulting equations are far too
complicated to be of any
use. There is no other possible coupling of the form (3.2) with the
$q$-P$_{\rm II}$ (3.23).

Let us now turn to the case of a linear coupling given by equation (3.3).
In the case of
d-P$_{\rm II}$ (3.22) we require that the only singularities of the
coupling terms $(\alpha
x + \beta)/(\gamma x + \delta)$ be the two singularities $\pm 1$. This
leads to a coupling
of the form:
$$
\overline y = {\alpha(x\pm 1)y +1 \over x\mp 1}, \eqno(3.27)
$$
where one of the parameters (e.g. $\gamma$) has been put to $1$ through the
appropriate gauge
of $y$. Computing the successive $y$'s we find that the condition for
having a finite
$\overline{\overline{\overline{y}}}$, depending on the initial condition
$y$, is just
$\overline
\alpha \alpha=1$. This means that all even $\alpha$'s are constant while
all odd ones are
equal to the inverse of this constant.

For $q$-P$_{\rm II}$ (3.23), in the case where (3.4) has $0$ and $1$ as
roots, we have two
 possible couplings:
$$
\overline y = {\alpha xy +1 \over x-1} \eqno(3.28a)
$$
and
$$
\overline y = {\alpha (x-1)y +1 \over x}. \eqno(3.28b)
$$
It turns out that in both cases the confinement condition is the same as in
the case of
d-P$_{\rm II}$ namely $\overline \alpha \alpha =1$. When the roots are $0$
and $q$, the
possible couplings are:
$$
\overline y = {\alpha xy +1 \over x-q} \eqno(3.29a)
$$
and
$$
\overline y = {\alpha (x-q)y +1 \over x}. \eqno(3.29b)
$$
The condition for integrability in the two cases is $\alpha=1/\lambda$. Two
other
couplings are possible when the roots of (3.4) are $q$ and $\infty$:
$$
\overline y = {\alpha y +1 \over x-q}, \eqno(3.30a)
$$
and
$$
\overline y= \alpha(x-q)y+1. \eqno(3.30b)
$$
The integrability condition for (3.30a)  is $\alpha=\underline{q}$ and for
(3.30b), it is
$\alpha=1/q$.
Finally if (3.4) has $1$ and $q$ as roots, the possible couplings are:
$$
\overline y = {\alpha(x-1)y+1 \over x-q} \eqno(3.31a)
$$
and
$$
\overline y = {\alpha (x-q)y +1 \over x-1}. \eqno(3.31b)
$$
In the case of (3.31a), the integrability condition is
$$
\overline \alpha \alpha \underline \alpha (-\overline{\overline q} + 1)
+\overline \alpha
(-\overline{q}q + \overline{\overline q}+\underline q - 1) + q^2 -
\overline q=0
\eqno(3.32)
$$
and in the case of (3.31b), the condition reads
$$
\overline \alpha \alpha \underline \alpha ({\overline q}^2 - q) +\overline
\alpha
\alpha (-\overline q q + \overline{\overline q}+\underline q -1) -
\underline q + 1=0.
\eqno(3.33)
$$
Equations (3.32) and (3.33) are integrable and they belong to the family of
linearisable equations
[10].

 One last remark is necessary at this point since we have seen that almost
all the equations we obtained contain terms with binary, ternary or
quaternary freedom. The
presence of these terms indicates that our systems must be augmented by
adding more
components. This will not alter the order of the resulting equation: it
just increases the
number of its parameters. The continuous limit is, of course, affected by
this choice.

\bigskip
\noindent {\scap 4. Conclusion}
\medskip
\noindent

In this work we have presented a systematic approach for the construction
of integrable
 third order systems through the coupling of a second order equation to a
Riccati or a
linear first order equation. Thus we have extended the Gambier approach
(first used in his
derivation of the second order ODE that bears his name) to higher order
systems. We have
applied this coupling method to both continuous and discrete systems (given
that we have
already presented [4] the discrete equivalent of the Gambier equation).

One point remains to be discussed. It is often argued that, since the
Riccati is a
linearisable equation, the coupling of the Riccati to another of the same
kind or to an
integrable second order is always integrable. The (na\"{\i}ve) argument is
the following:
first solve the first equation, substitute the solution into the second and
solve it by
linearising it. The argument about singularities is usually swept aside by
the statement that
one is interested only in solutions on the real-time axis. However the
situation is not that
simple. What integrability consists in is a global description of the
solutions of the
equations. The argument about solutions on the real-time axis is not
acceptable since it
offers just a local description of the solution of the equation. A global
representation of the
solution of a linear equation (and, thus, also of a Riccati) involves path
integrals winding
over the complex-time plane. Thus the study of movable singularities is
crucial and the
Painlev\'e property a necessary condition for integrability of the systems.

How do these
arguments carry over to the discrete setting? One must go back to the way
difference equations
are formally solved. Given a linear difference (or $q-$) equation, we can
express the solution
as an infinite product of matrices, the elements of which  depend on the
coefficients of the
equation. A singularity appears whenever one of the matrices is singular.
In this case the
solution of the linear difference equation cannot be defined for every $n$.
However it is
in general possible to choose the coefficients of the equation so as to
avoid these
singularities. In the case of a coupling the coefficients depend on the
solutions of some
other equation. Thus there is no way to control the singularities (which
depend on the
initial conditions of the first equation). As a consequence the solution of
the second
equation cannot be defined everywhere unless the confinement property is
satisfied. Thus,
again, despite the linearisability of the discrete Riccati, whenever we
talk about a global
description of the solution of the coupled system, the application of the
adequate
integrability criterion is mandatory.

\bigskip
\noindent {\scap Acknowledgments}
\medskip
\noindent S. Lafortune acknowledges two scholarships: one from NSERC
(National Science and
Engineering Research Council of Canada) for his Ph.D. and one from ``Le
Programme de
Soutien de Cotutelle de Th\`ese de doctorat du Gouvernement du Qu\'ebec''
for his stay
in Paris.

\bigskip
\noindent {\scap References}.
\bigskip
\item{[1]} P. Painlev\'e, Acta Math. 25 (1902) 1.
\item{[2]} B. Gambier, Acta Math. 33 (1910) 1.
\item{[3]} E.L. Ince, {\sl Ordinary differential equations}, Dover, New
York, 1956.
\item{[4]} B. Grammaticos and A. Ramani, Physica A 223 (1995) 125.
\item{[5]} J. Chazy, Acta Math. 34 (1910) 317.
\item{[6]} C.M. Cosgrove, {\sl Corrections and annotations to Ince's
chapter 14}.
\item{[7]} A. Ramani, B. Dorizzi, B. Grammaticos, Phys. Rev. Lett. 49
(1982) 1539.
\item{[8]} B. Grammaticos, A. Ramani and V. Papageorgiou, Phys. Rev. Lett.
67 (1991) 1825.
\item{[9]} B. Grammaticos, A. Ramani and P. Winternitz, {\sl Dicretizing
families of
 linearizable equations}, preprint (1997).
\item{[10]} A. Ramani, B. Grammaticos and G. Karra, Physica A181 (1992) 115.
\item{[11]}	A. Ramani, B. Grammaticos, Physica A 228 (1996) 160.
\end